\newcommand{\1}{d} 
\newcommand{\2}{u} 
\newcommand{\3}{s}

\newcommand{\la}[1]{\label{#1}}
\newcommand{\be}{\begin{equation}}
\newcommand{\ee}{\end{equation}}
\newcommand{\ba}{\begin{eqnarray}}
\newcommand{\ea}{\end{eqnarray}}
\newcommand{\bi}{\begin{itemize}}
\newcommand{\ei}{\end{itemize}}

\newcommand{\tr}{{\rm Tr\,}}

\newcommand{\Hc}{{\rm H.c.\ }}

\newcommand{\fr}[2]{{\frac{#1}{#2}}}

\newcommand{\Nf}{N_{\rm f}}

\documentclass{PoS}

\title{Towards a quantitative understanding of the $\Delta I=1/2$ rule}

\ShortTitle{The $\Delta I=1/2$ rule}

\author{\speaker{P.~Hern\'andez}\thanks{The work reviewed here has been done in collaboration with L.~Giusti, M.~Laine, M.~L\"uscher, C.~Pena, P.~Weisz, J.~Wennekers and H.~Wittig.}\\\\
        IFIC and Universidad de Valencia, Edif. Institutos Investigaci\'on Apt 22085, 46071 Valencia, Spain\\
        E-mail: \email{pilar.hernandez@ific.uv.es}}

\abstract{A recently proposed strategy to quantify the role of the charm quark mass in the 
 $\Delta I=1/2$ rule is reviewed. Results for the low-energy couplings of the $\Delta S=1$ chiral effective Hamiltonian in a theory with a light charm quark
(GIM limit) are obtained through a matching of three-point correlation functions computed in Chiral Perturbation Theory and in quenched lattice QCD. We observe a large $\Delta I=1/2$ enhancement, which is not large enough to explain the experimental result, but suggests significant long-distance contributions to the physical $\Delta I=1/2$ which are unrelated to penguin contractions or operators. }

\FullConference{XXIV International Symposium on Lattice Field Theory\\
		 July 23-28 2006\\
		 Tucson Arizona, US}

\begin{document}

\section{The $\Delta I=1/2$ rule}

One of the most striking hierarchies in hadronic physics is the famous $\Delta I=1/2$ rule, which refers to the experimental observation that the kaon decay 
amplitude in two pions with total isospin, $I=0$, is twenty times larger than that into an $I=2$ state:
%\ba
%T(K^+ \rightarrow \pi^+ \pi^0) &=& {\sqrt{3}\over 2} A_2 e^{i \delta_2}  \nonumber\\
%T(K^0 \rightarrow \pi^+ \pi^-) &=& {1\over\sqrt{6}} A_2 e^{i \delta_2} + {1\over \sqrt{3}} A_0 e^{i \delta_0}    \nonumber\\
%T(K^0 \rightarrow \pi^0 \pi^0) &=& {\sqrt{2}\over \sqrt{3}} A_2 e^{i \delta_2} - {1\over \sqrt{3}} A_0 e^{i \delta_0}, 
%\ea
% where 
\ba
 T\left(K^0\rightarrow \pi\pi|_{I=\alpha}\right) = A_\alpha e^{i \delta_\alpha}
~~~~~{A_0 \over A_2} = 22.1.
\ea

The explanation of this large number is one of the notorious failures of large $N_C$ which predicts~\cite{largenc} the ratio to be $\sqrt{2}$.
 
It was soon realized that there are many scales relevant in the dynamics of these decays, $M_W, m_c, M_K,...$  and maybe subleading orders in $N_C$ could get enhanced by large renormalization group logarithms \cite{oldies}. 

Below the scale of the $W$ mass, this boson can be integrated out of the theory. The resulting effective Hamiltonian for the $\Delta S=1$ transitions can be obtained through the Operator Product Expansion (OPE) of two left-handed currents. The use of CPS symmetry and the flavour symmetry restricts the possible dimension six four-quark operators to just two, $Q_1^{\pm}$ and 
$Q_2^{\pm}$, which are singlets under the $SU(4)_R$ and transform in 
the $\mathbf{84}$ and $\mathbf{20}$ representations of $SU(4)_L$ \cite{georgi}:
\ba
{\cal{H}}^{QCD}_{\rm w}= 2\sqrt{2} G_F (V_{us})^*V_{ud}
  \sum_{\sigma=\pm}
       k_1^{\sigma}Q_1^{\sigma}  + k_2^{\sigma}Q_2^{\sigma}, ~~~
\label{eq:su4heff}
\ea
where
\ba
& & Q_1^{\pm} = \Big\{
   (\bar{s}\gamma_{\mu}P_{-}{u})({\bar u}\gamma_{\mu}P_{-}{d})
\pm(\bar{s}\gamma_{\mu}P_{-}{d})({\bar u}\gamma_{\mu}P_{-}{u})
   \Big\} - (u\,{\to}\,c), \nonumber \\
& & Q_2^{\pm} =\left(m_u^2 -m_c^2\right) \Big\{ m_d \left(\bar{s}P_+ d\right) + m_s \left(\bar{s} P_- d\right)\Big\}.
\ea
The operators $Q_2^{\pm}$ do not contribute to the physical amplitudes and vanish identically if $m_u=m_c$.  
The $\mathbf{ 84}$ operator, $Q_1^+$, contributes both to the $\Delta I=1/2, 3/2$ transitions, while the $\mathbf{20}$, $Q_1^-$, only contributes to $\Delta I=1/2$. 
This Hamiltonian can be run down to lower scales resumming leading logarithms through the renormalization group. A $\Delta I=1/2$ enhancement of the Wilson coefficients is then 
observed at low scales in commonly used schemes~\cite{oldies}. For example in the renormalization group invariant scheme (RGI) at 2-loops:
\ba
\left[{k_1^-(\mu)\over k_1^+(\mu)}\right]_{\mu=m_c} \simeq 2.
\label{eq:short}
\ea
Although this goes in the right direction, a much bigger hierarchy must come from the hadronic matrix elements to match the experiment. Besides at $\mu=m_c$ higher-order corrections to eq.~(\ref{eq:short}) are ${\cal O}$(100$\%$).

One can nevertheless try to push this perturbative analysis 
below the charm quark mass to obtain some qualitative understanding. When the charm is integrated out of the theory, one moves from a four-flavour theory to a three-flavour one. The OPE then allows any four-quark operator that transforms as the $\mathbf{27}$ or $\mathbf{8}$ dimensional representations of $SU(3)_L$. Among the latter there are operators that are no longer the product of two left currents, but 
involve the product of left and right currents or densities, such as the famous {\it penguin} operators. Arguments were put forward to argue that the hadronic matrix elements of the penguin operators are larger than those of the left current operators and this could be the origin of the rule \cite{vzs}. 

The computation of these matrix elements requires however a non-perturbative method. Many approaches to estimate the matrix elements using large $N_C$ arguments or models have been pursued in the past.
For a review and further references see e.g. Ref.~\cite{reviews}. Although plausible arguments seem to indicate that the rule could be the result of the accumulation of several instances of the octect enhacement, combined with large penguin matrix elements \cite{bbg}, the approach relies on the use of perturbation theory down to dangerously low scales. 

\section{The $\Delta I=1/2$ rule on the lattice}

It was realized in the early days of Lattice QCD \cite{early} that the $\Delta I=1/2$ rule would be a very well suited problem for the lattice approach, 
since 
it is such a large effect! Even if there are approximations like  
quenching, or in the presence of large systematic uncertainties, such a large enhancement would be hard to miss.

In the pioneering work of Ref.~\cite{bernardetal}, it was proposed to use the lattice to perform the matching of the $\Delta S=1$ effective Hamiltonian of eq.~(\ref{eq:su4heff}) to an effective Hamiltonian in terms of the hadronic degrees of freedom, that is a chiral Lagrangian. The possibility to include the $\Delta S=1$ 
interactions in the chiral Lagrangian as a perturbation was first proposed 
by Cronin in Ref.~\cite{cronin}. In addition to the chiral Lagrangian that describes the strong interactions, ${\cal L}_{\chi PT}$, one would have  
an effective weak Hamiltonian with the same flavour symmetries as that in eq.~(\ref{eq:su4heff}), ${\cal H}^\chi_w$. The operators $Q_i^\pm$ can be decomposed into $\mathbf{27}$ and  $\mathbf{8}$ of $SU(3)_L$. Therefore all the operators that can be constructed with the building blocks of the chiral Lagrangian, $U, \partial_\mu, M$, with the same transformation properties as $Q_i^\pm$ should be included. 
At the leading order (LO) in a momentum or mass expansion there are just three of them, one 27-plet and two octects. The weak Hamiltonian can be writen \cite{bernardetal}:
\be
  {\cal H}^\chi_w \equiv  2 \sqrt{2} G_F V_{ud} V^*_{us}
  \biggl\{ 
  \frac{5}{3} g_{27} {\cal O}_{27} 
  + 2 g_8 {\cal O}_8
  + 2 g_8'{\cal O}'_8
  \biggr\} + \Hc   \; ,
 \la{Lw_XPT}
\ee
where $g_8, g_{27}$ and $g_8'$ are low-energy couplings that contain the non-perturbative dynamics that is not fixed by symmetry arguments. 
The operators read
\ba
 {\cal O}_{27} & \equiv & 
 {[\; \hat {\! {\cal O}}_{w}]}^+_{\3\2\2\1} = 
 \fr35\Bigl(  
 {[ {\cal O}_{w} ]}_{\3\2\1\2} + 
 \fr23 {[  {\cal O}_{w} ]}_{\3\2\2\1}
 \Bigr)
 \;, \la{formofO} \\   
 {[{\cal O}_{w}]}_{rsuv} & \equiv &  
 \frac{F^4}{4} 
 \Bigl(\partial_\mu U U^\dagger\Bigr)_{ur}
 \Bigl(\partial_\mu U U^\dagger\Bigr)_{vs}
 \;, \la{O_XPT} \\
 {\cal O}_{8} & \equiv &   \fr12 \sum_{k=u,d,s}
 {[ {\cal O}_{w} ]}_{\3 k k\1}
 \;, \la{formofR} \\
 {\cal O}'_{8} & \equiv & 
  \frac{F^2}{2} {\Sigma}
  \Bigl( e^{i\theta/\Nf} M U 
   + U^\dagger M^\dagger e^{-i\theta/\Nf} \Bigr)_{\1\3} 
   \;, \la{formofO8p}
\ea
where we have made use of $\tr[\partial_\mu U U^\dagger] = 0$
to simplify the expressions.

If these couplings were known,  the ratio of the 
$\Delta S=1$ amplitudes at LO in the chiral expansion would be:
\ba
{A_0 \over A_2} = {1 \over \sqrt{2}} \left({1\over 5} + {9\over 5} {g_{8} \over g_{27}}\right).
\ea
LO chiral perturbation theory is not expected to be very precise at the scale of $M_K$, but again the enhancement is such a large effect that as long as 
the expansion is reasonably well-behaved, the effect should already be there at the LO. 

It was then proposed in \cite{bernardetal} that these low-energy couplings  could be determined by matching appropriate (the simplest) correlation functions between the chiral effective theory and lattice QCD, and in particular that this could be done through the computation of 
three-point functions and two-point functions. Note that to compute directly the amplitude $K \rightarrow \pi\pi$, four-point functions would be needed. 

This nice proposal turned out to be extremely difficult to implement in practice. Firstly the renormalization of four-fermion operators  is extremely challenging when there is explicit breaking of chiral symmetry in the regularization. Not only there are a large number of additional mixings with wrong-chirality operators, but there is even mixing with lower dimensional ones and therefore power-diverging coefficients \cite{romans}. With the advent of Ginsparg-Wilson regularizations \cite{gw,perfect,overlap}, that preserve an exact chiral symmetry 
\cite{exact}, the renormalization of the effective Hamiltonian of eq.~(\ref{eq:su4heff}) becomes as simple as it is in the continuum \cite{rengw}\footnote{Only in Ref.~\cite{strat} the mixing with GIM suppressed operators $Q_2^\pm$ has been discussed.}:
\ba
 & & {Q}_1^{\pm} =
     {Z}_{11}^{\pm}{Q}_1^{\pm,\rm{bare}}
    +{Z}_{12}^{\pm}{Q}_2^{\pm,\rm{bare}}, \nonumber\\
 & & {Q}_2^{\pm} = {Z}_{22}^{\pm}{Q}_2^{\pm,\rm{bare}}.
\ea
Indeed the computations of the $\Delta I=1/2$ rule in Refs.~\cite{rbc,cppacs} have been performed in the quenched approximation 
using domain-wall
fermions \cite{dw}, that approximately satisfy the Ginsparg-Wilson
relation. The lattice spacing used in these computations was 
however too low to keep the charm active, so they considered the
effective Hamiltonian where the charm is integrated out perturbatively. The effect of integrating out the charm brings in important complications. On the one hand, the renormalization involves power-divergent subtractions, which require a very good control over statistical as well as systematic uncertainties. 
A second difficulty was pointed out in Ref.~\cite{gp} and is related to 
the quenched ambiguities, which occur at the level of the OPE. More concretely there are operators in the OPE that are octects under the valence group and contain ghost quarks, in the supersymmetric formulation of the quenched approximation \cite{susy}.  Now, if these operators are not included in the OPE, by assuming for instance that the quenched approximation is only used to define the matrix elements at some low scale, it is unclear what the meaning of non-perturbative renormalization is, since it requires to compute the matrix elements in the quenched approximation up to very high scales. In other words it is not clear whether the use of the quenched approximation to compute matrix elements combined with the full theory OPE is really a consistent method.

%It is well-known that in general the strong 
%chiral effective
%theory in the quenched approximation contains spurious couplings, because of the presence of the $U(1)_A$ field, which does not decouple in this approximation. In the case of the weak Hamiltonian, spurious couplings can also appear. Some are related to the $U(1)_A$ field, but others are related 
%to the fact that the number of valence, $N_v$, and sea quarks, $N_s$, is different. In the replica method to define the quenched approximation \cite{replica},  the spurious operators can always be chosen in such a way that they vanish in the limit $N_v \rightarrow N_s$ \cite{HerLai3}.

%A resasonable working assumption in this case would be to assume that the couplings associated with the same operators that appear in the full theory, and do not vanish in the limit $N_v \rightarrow N_s$, are the ones that should be closer to those in the full theory. 
%At this point the complication would just ammount to the need to determine more couplings simultaneously. 

%It turns out however that when the charm is integrated out the same ambiguities that appear in the chiral effective theory affect already the definition of the effective Hamiltonian at the quark level. 

Finally the simulations in Refs.~\cite{rbc,cppacs} were carried out at relatively large quark masses above $m_s/2$. Chiral corrections were shown to be very large and a large systematic error resulted from the long chiral extrapolations. 
In the end, both collaborations found a large enhancement but there was almost a factor of 2 discrepancy between the two computations. 
  
\section{New strategy}
   
The approach to the $\Delta I=1/2$ rule that was presented in Ref.~\cite{strat} was 
designed not to reach the final result directly, but to try to reveal if the large enhancement is coming from one single leading 
effect. The point is that on the lattice, 
the different physical scales that are involved in these decays can be modified at will in order to 
understand their relevance. In particular this is quite clearly the case for the charm quark mass. 

Most of the ideas that have been put forward to explain the enhancement are related in one way or the other to the charm quark mass. If the large enhancement is due to the large separation between $m_c$ and $\Lambda_{QCD}$ or the up quark mass, there should be no enhancement whatsoever in a theory with a light charm quark.

More concretely the strategy that we proposed to quantify the role of the charm quark mass is the following:
\begin{itemize}
\item Step 1: light charm quark 

A theory with four degenerate quarks, $m_c=m_u=m_d=m_s$ (GIM limit) is matched to a
$SU(4)$ chiral effective theory to extract the low-energy couplings that would mediate $\Delta S=1$ transitions

\item Step 2: $\Lambda_{\chi PT} \gg m_c \gg m_u = m_d = m_s$

If the charm gets significantly heavier than the up quark, but remains in the realm of the chiral theory, that is charmed mesons are well below the chiral theory cutoff $4\pi F$, the charm quark can be integrated out from the $SU(4)$ chiral theory to obtain the $SU(3)$ chiral effective theory. If the charm quark is not too large this can be done analytically in chiral perturbation theory \cite{HerLai2}. 

\item Step 3: $m_c \geq \Lambda_{\chi PT} \gg m_u=m_d=m_s$

If the charm gets too heavy to be describable in terms of an effective chiral theory, the matching to the $SU(3)$ chiral theory has to be done non-perturbatively. The couplings $g_{27}(m_c)$ and $g_8(m_c)$ can then be monitored as a function of the charm quark mass.

\end{itemize}

\section{$K\rightarrow\pi\pi$ amplitudes in the GIM limit}

We will now describe the formulation of the problem in the GIM limit, that is for $m_u=m_d=m_s=m_c$. 

\subsection{Lattice formulation}

In this limit there are only two operators in the OPE at first order in $G_F$, $Q_1^{\pm}$ in eq.~(\ref{eq:su4heff}), which renormalize therefore
multiplicatively
\ba
Q_1^{\pm} = {Z}_{11}^{\pm} Q_1^{\pm,\rm{bare}}.
\ea
In the quenched approximation no spurious operator can appear \cite{strat}.

It can be shown \cite{strat} that using overlap fermions \cite{overlap,exact,locality}  this simple renormalization pattern is preserved.
In Ref.~\cite{renorm}, the renormalization constants have been computed non-perturbatively through an intermediate matching to twisted-mass Wilson fermions at some large reference quark mass. The corresponding renormalization constants for twisted-mass Wilson fermions have been previously computed
using the Schr\"odinger functional approach \cite{tmWilson}. 

The result for the renormalization constants in the RGI scheme (for details see \cite{strat,renorm}) is summarized in Table~\ref{tab:ren}, where the perturbative (one-loop) and mean-field estimates are also shown for comparison. 
\begin{table}\begin{center}
\begin{tabular}{c c c c}
\hline\hline
\noalign{\vskip0.5ex}
    & bare P.T. & MFI P.T. & N.P. \\
\noalign{\vskip0.5ex}
\hline
\noalign{\vskip0.3ex}
$Z_{11}^{-}/Z_{11}^{+}$ & 0.525 & 0.582 & 0.584(62) \\
$Z_{11}^{+}/Z_{\rm A}^2$     & 1.242 & 1.193 & 1.15(12) \\
$Z_{11}^{-}/Z_{\rm A}^2$     & 0.657 & 0.705 & 0.561(61)\\
\noalign{\vskip0.3ex}
\hline\hline
\end{tabular}
\caption{Comparison of the perturbative, mean-field improved and non-perturbative renormalization constants in the RGI scheme. }
\label{tab:ren}
\end{center}
\end{table}

\subsection{$\chi PT$ formulation}

In the GIM limit, the chiral Lagrangian has a $SU(4)_L \times SU(4)_R$ symetry group. It can be shown that at leading order in the momentum expansion, and in contrast with the $SU(3)$ case, only two operators appear 
one $\mathbf{84}$ and a $\mathbf{20}$:
\ba
   {\cal H}_{\rm w}^{\rm \chi PT
} =
   2\sqrt{2} G_F~(V_{us})^*V_{ud}
   \sum_{\sigma=\pm}g^\sigma  [{\cal O}^\sigma]\nonumber
\label{eq_HwChPT}
\ea
where
\ba
  {\cal O}^{\pm} =
  {F^4\over 4}\left[
   \left(U\partial_\mu U^\dagger\right)_{us}
  \left(U\partial_\mu U^\dagger\right)_{du}\pm \left(U\partial_\mu U^\dagger\right)_{uu}
  \left(U\partial_\mu U^\dagger\right)_{ds} - (u\rightarrow c)\right]. \nonumber
\label{eq_O1_ChPT}
\ea
The normalization is such that in the large $N_C$ limit
\ba
[g^+]_{N_c} = [g^-]_{N_c} = 1. 
\ea
The ratio of the $K\rightarrow \pi\pi$ amplitudes in the GIM limit would  be given by
\ba
   \frac{A_0}{A_2} = \frac{1}{\sqrt{2}}\left(\frac{1}{2}+\frac{3}{2}\frac{g^{-}}{g^{+}} \right), \;\;\;\;\;\; 
\ea
and therefore at LO in the chiral expansion the hierarchy of the amplitudes is directly related to a hierarchy in the couplings
$g_+$ and $g_-$. Our primary goal is therefore to determine these couplings. 
 
We also note that no further operators appear in the quenched or partially quenched approximation at this order, so there are
no Golterman-Pallante ambiguities in the effective theory~\cite{strat}. 

\subsection{The matching}

In order to determine the couplings $g^{\pm}$, we will match suitable correlation functions computed in lattice QCD to those computed to next-to-leading (NLO) order in the chiral expansion. Even though we will do the matching at very small quark masses, we will be in a finite volume, entering the $\epsilon$-regime of chiral perturbation theory \cite{epsilon}. 

We will consider three-point bare correlation functions of the four-fermion operators and two left-handed currents \cite{ghlw}:
\ba
C^\sigma(x_0,y_0) \equiv \sum_{\mathbf{x},\mathbf{y}} \left\langle [J_{L0}(x)]_{du}\,Q^\pm(0)\,[J_{L0}(y)]_{us}\right\rangle .
\label{eq:3pt}
\ea
More concretely we will match the ratio of these three-point functions to left-current correlators:
\ba
R^{\sigma}(x_0,y_0) \equiv  \frac{C^\sigma(x_0,y_0)}{C(x_0)C(y_0)}, 
\ea
where 
\ba
C(x_0) \equiv \sum_{\mathbf{x}} \left\langle [J_{L0}(x)]_{\alpha\beta}[J_{L0}(0)]_{\beta\alpha}\right\rangle .
\ea

Factorizing out the unknown couplings, the same ratio can be computed in the effective theory \cite{HerLai1}:
\ba
{\cal R}^{\sigma}(x_0,y_0) \equiv \frac{{\cal C}^\sigma(x_0,y_0)}{{\cal C}(x_0){\cal C}(y_0)}, 
\ea
where
\ba
{\cal C}^\pm(x_0,y_0) &\equiv & \int d^3x\int d^3y \left\langle [{\cal J}_{L0}(x)]_{du}\,{\cal O}^\pm(0)\,[{\cal J}_{L0}(y)]_{us}\right\rangle \\
{\cal C}^\pm(x_0) &\equiv & \int d^3 x \left\langle [{\cal J}_{L0}(x)]_{\alpha\beta}[{\cal J}_{L0}(0)]_{\beta\alpha}\right\rangle .
\label{eq:chptcorrs}
\ea 
The couplings can the be extracted from the matching:
\ba
g^\pm  = k^{\sigma}(M_W) {Z_{11}^{\sigma}(g_0)\over Z_A^2} {R^\sigma(x_0,y_0) \over {\cal R}^\sigma(x_0,y_0)} ,
\ea
where the Wilson coefficients are obtained in the same renormalization scheme as the operators, i.e. the RGI one. For the explicit expressions 
see \cite{strat}.

\subsection{${\cal R}^\sigma$ to NLO in $\chi PT$}

In order to obtain the LO couplings, the matching between lattice QCD and $\chi$PT should be done as close as possible to the chiral limit. 
The use of Ginsparg-Wilson regularization makes it possible to do simulations with extremely small quark masses: below a few MeV with 
volumes larger than $2$~fm. However it becomes very costly to increase the volume much further than $2$~fm or so. While the usual way to take the chiral limit is to first take $V$ to infinity and only then take $m$ to zero, there are advantages in taking $m$ as small as possible at finite $V$, entering the so-called $\epsilon$-regime defined by the condition:
\ba
m \Sigma V \leq 1.
\label{eq:epsilon}
\ea
For $F L \gg 1$, this limit is equivalent to $M L \ll 1$, so the Compton wavelength of the pion is larger than the box size. Finite volume effects are large in this regime, however they are calculable in chiral perturbation theory \cite{epsilon}. The counting rules of the chiral expansion that ensure the condition of eq.~(\ref{eq:epsilon}) are:
\ba
m \Sigma \sim \epsilon^4 ~~~~~L^{-1}, T^{-1} \sim \epsilon ~~~~ p \sim \epsilon
\ea
which are different to the usual ones. As a result of the new power-counting: 
\begin{itemize}
\item The zero-momentum modes of the pions become non-perturbative and have to be resumed to all orders. This is achieved by factorizing out the constant field configurations and treating them as collective variables in the partition functional.
%:
%\ba
%U = U_0 U_\xi = U_0\,e^{\,i\,2 \xi(x)/F}\;\;\; \int dx \xi(x) = 0 \\
%{\mathcal Z}= {\int_{{ SU(N_f)}}}
%dU_o\, \int d\xi\;\; e^{-\mathcal{S}_\chi(U_o,\xi)}.
%\ea
\item There is a reordering of the chiral expansion and at any given order less relevant couplings appear. In particular for the processes at hand
 the strong or weak higher-order couplings \cite{kamboretal,HerLai3} shown in Table~\ref{tab:nlo} that would appear at NLO in infinite (or large enough) volume are all suppressed at the same order 
in the $\epsilon$-regime.
\end{itemize}
\begin{table}
\begin{center}
\begin{tabular}{cll}
 &  $p$-regime & $\epsilon$-regime \\\\\hline\\
 ${\cal L}_{QCD}$&  $L_4 \,\langle D_\mu U^\dagger D^\mu U\rangle\,
   \langle {\cal S} \rangle$  & $\times$ \\
 & $L_5 \,\langle D_\mu U^\dagger D^\mu U ~{\cal S} \rangle\,$ & $\times$ \\
 & $L_6 \,\langle {\cal S} \rangle^2$ &  $\times$ \\ 
 & $L_8 \,\langle {\cal S}^2 \rangle$ & $\times$ \\
\\\\\hline\\
${\cal H}^{\chi PT}_{w}$& $D^{\pm}_{2} t^{\pm}_{ij,kl} \, 
{\cal P}_{ij} {\cal P}_{kl} \,$& $\times$ \\
 & $D^{\pm}_{4} t^{\pm}_{ij,kl} \, 
\left({\cal L}_\mu\right)_{ij} \left(\left\{ {\cal L}^\mu, {\cal S}\right\} \right)_{kl} \,$ & $\times$ \\
 & $D^{\pm}_{7} t^{\pm}_{ij,kl} \, \left({\cal L}_\mu \right)_{ij} 
\left({\cal L}_\mu \right)_{kl} \langle {\cal S}\rangle \,$& $\times$ \\
 & $D^{\pm}_{20}  t^{\pm}_{ij,kl} \, 
\left({\cal  L}_{\mu} \right)_{ij}
\left(\partial_\nu {\cal W}_{\mu\nu}\right)_{kl}\,$ & $\times$ \\
 & $D^{\pm}_{24}  t^{\pm}_{ij,kl} \, 
\left({\cal  W}_{\mu\nu} \right)_{ij} 
\left({\cal W}_{\mu\nu}\right)_{kl}\,$ & $\times$ \\
\\\\\hline\\
\end{tabular}
\caption{Strong and weak operators that would contribute to the observables considered in eqs.(\protect\ref{eq:chptcorrs}) at NLO in infinite-volume chiral perturbation theory. ${\cal S}\equiv U\chi^\dagger + \chi U^\dagger$, ${\cal P}\equiv i(U\chi^\dagger - \chi U^\dagger)$, ${\cal L}_\mu \equiv U \partial_\mu U^\dagger$and ${\cal W}_{\mu\nu}\equiv 2 (\partial_\mu {\cal L}_\nu + \partial_\nu {\cal L}_\mu)$, where $\chi \equiv 2 m \Sigma/F^2$. The tensors $t^\pm_{ij,kl}$ project onto the appropriate representations $\mathbf{84}$ or $\mathbf{20}$.}
\label{tab:nlo}
\end{center}
\end{table}

An additional complication is the quenched approximation, where we shall be working. The quenched chiral-perturbation-theory version of the $\epsilon$-regime was studied in detail in Ref.~\cite{epsilonquenched}. Analytical treatment is possible if averages are considered in fixed topological sectors. Ginsparg-Wilson fermions satisfy an exact index theorem \cite{perfect} and therefore these averages can be computed on the lattice aswell. 
The different role of topology in the $\epsilon$ and $p$-regimes was first discused in \cite{leutsmil}. The fixing of topology can be seen as a type of boundary condition, which should not affect the local properties of the theory, and therefore the dependence on the topological charge is in principle predictable in terms of the low-energy couplings of the effective theory, just as the finite-volume dependence is. 

\begin{figure}
\begin{center}
%%%%%%%%%%%%%%%%%%%%%%%%%%%%%%%% FIGURE %%%%%%%%%%%%%%%%%%%%%%%%%%%%%%%%%%%%
\includegraphics[width=.4\textwidth]{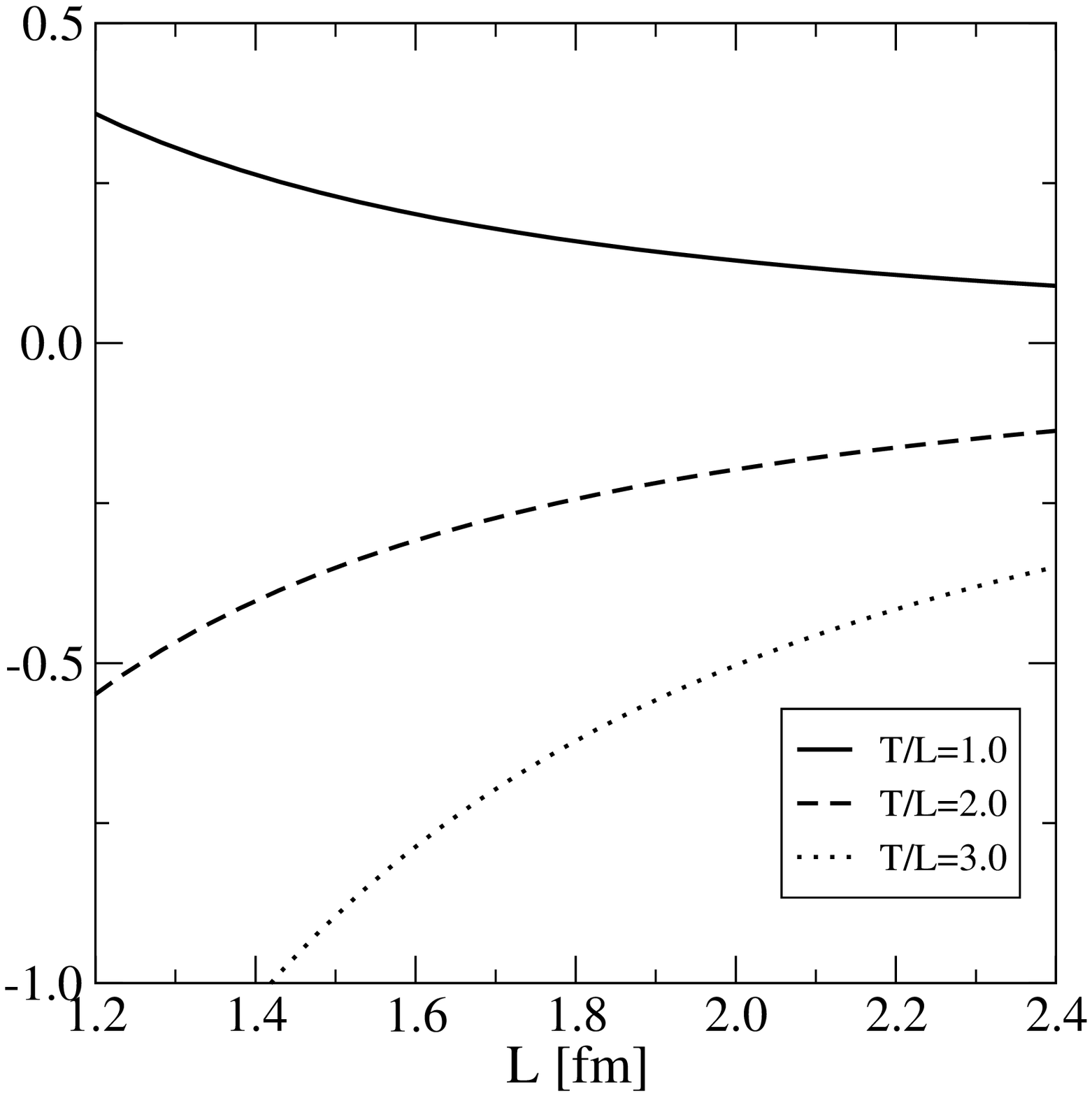} \includegraphics[width=.42\textwidth]{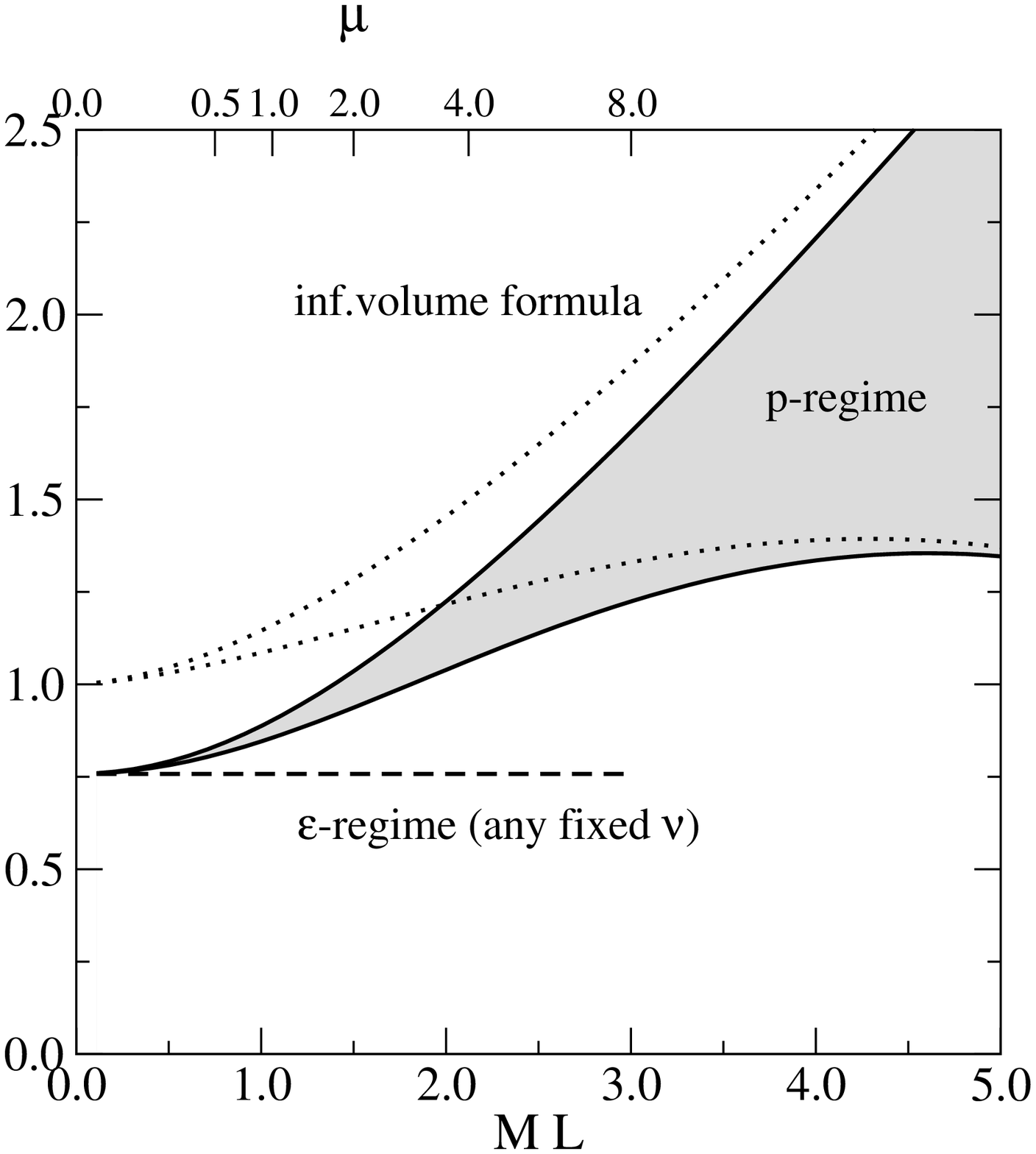}
\caption{Left: NLO correction $K$ as defined in eq.~(\protect\ref{eq:k}) as a function of the spatial extent of the box $L$ in fm for three different ratios $T/L$. Right: $2 R^+(-T/3,T/3)$ for a lattice of 2~fm as a function of $M L$ for $T/L =2$. The band corresponds to varying the NLO couplings within a large reasonable range. The dotted and dashed lines are the $\infty$-volume and the $\epsilon$-regime results respectively.}
\label{fig:r}
%%%%%%%%%%%%%%%%%%%%%%%%%%%%%%%%%%%%%%%%%%%%%%%%%%%%%%%%%%%%%%%%%%%%%%%%%%%%
\end{center}
\end{figure}

The result of ${\cal R}_\nu^\sigma$ at NLO in the $\epsilon$-regime, in a fixed-topological sector of charge $\nu$, was computed in \cite{HerLai1,strat} and the result is:
\ba
2~ {\cal R}_\nu^\pm(x_0,y_0)= 
1 \pm {2 \over (F L)^2} \left[ \rho^{-1/2} \beta_1 - \rho k_{00}\right] = 1 \pm K ,
\label{eq:k}
\ea
with $\rho \equiv T/L$ and $\beta_1, k_{00}$ are shape coefficients of the box  that depend only on $\rho$. 
The ratio turns out to be the same in the full and in the (partially-)quenched theories.   
It is quite remarkable that the result does not depend on the insertion of the currents $x_0,y_0$, 
it is the same for all topological sectors $\nu$, and as expected no higher order coupling enters at this order. 
In the left plot of Figure~\ref{fig:r} we show the numerical result for the NLO correction, $K$, as a function of the spatial extent of the box, 
for three values of the aspect ratio. If $\rho$ is not too large \footnote{The appropriate regime for large values of $\rho$ is the so-called $\delta$-regime, so actually the $\epsilon$-regime 
expansion breaks down in the $\rho \rightarrow \infty$ limit. }, the corrections are reasonably small for lattice sizes above $2$~fm or so. 

Since we will simulate a number of quark masses, including those for
 which $m \Sigma V \gg 1$, which is the usual regime of all previous
 calculations, we would need the results for the ratio ${\cal
 R}^\sigma$ in the $p$-regime, where the counting rules are the same
 as in infinite volume and therefore at NLO the couplings of
 Table~\ref{tab:nlo} become relevant. The result for ${\cal R}^\sigma$ in the $p$-regime has been
 presented in \cite{HerLai3} and is shown in the right plot of
 Figure~\ref{fig:r} as a function of $M L$ for a lattice of extent
 $2$~fm. The band corresponds to changing the value of the unknown NLO couplings within a resonable range. For comparison the $\epsilon$-regime and $\infty$-volume
 results are also shown. Surprisingly we find that deviations from the
 $\infty$-volume expectations are significant up to $M L \leq 5$ for
 these observables.

\subsection{$R^\sigma$ in lattice QCD}

The computation of $R^\sigma$ for small quark masses in a finite volume is non-trivial. It was observed that large fluctuations in 
observables containing quark propagators occur in the kinematical conditions of the $\epsilon$-regime \cite{fluctuations}. These fluctuations result from
the fact that the low-lying spectrum of the Dirac operator is discrete in this regime. Both the low-lying eigenvalues of the massless Dirac operator and the splittings between them are controlled by the 
quantity $(\Sigma V)^{-1}$:
\ba
\langle \lambda_i \rangle  \simeq {{\cal O}(1) \over \Sigma V} \qquad   \Delta \lambda = \lambda_{i+1} -\lambda_i \simeq {{\cal O}(1) \over \Sigma V},
\ea
therefore the eigenvalues of the massive Dirac operator in the $\epsilon$-regime where $m \leq (\Sigma V)^{-1}$ are of the same order as their splittings. In this situation, observables with 
point-to-all quark propagators can get large contributions from a few eigenfunctions. Space-time fluctuations in these eigenfunctions can lead to large and rare fluctuations in observables 
whenever peaks in the wavefunction happen to be near the fixed-point of the point-to-all propagator~\cite{2ptlma}. It has been shown~\cite{2ptlma} that these fluctuations could be cured provided 
$m \Sigma V \simeq 1$ through the technique of low-mode averaging (LMA) \cite{2ptlma,oldlma}. The idea is to rewrite the quark propagator as a sum over its spectral decomposition onto the subspace
spanned by a few low-lying eigenvalues (low part) and the rest (heavy part). Once this representation is included in the correlation functions, those contributions that 
contain the low-parts only, can be averaged over all possible spatial insertions of the sources, because the wavefunctions of the low-modes are known in all points. The mixed contributions often can also be 
averaged provided a few additional inversions of the Dirac operator on the low-modes are performed. This is indeed the case for the two-point functions \cite{2ptlma} and three-point functions \cite{strat} that
we need in the GIM limit. The contributions from just the heavy parts remain unaveraged, but since the low-modes are no longer affecting these contributions, they
should be much better behaved. 

\begin{figure}
\begin{center}
%\begin{tabular}{cc}
\includegraphics[height=0.3\textheight,angle=-90]{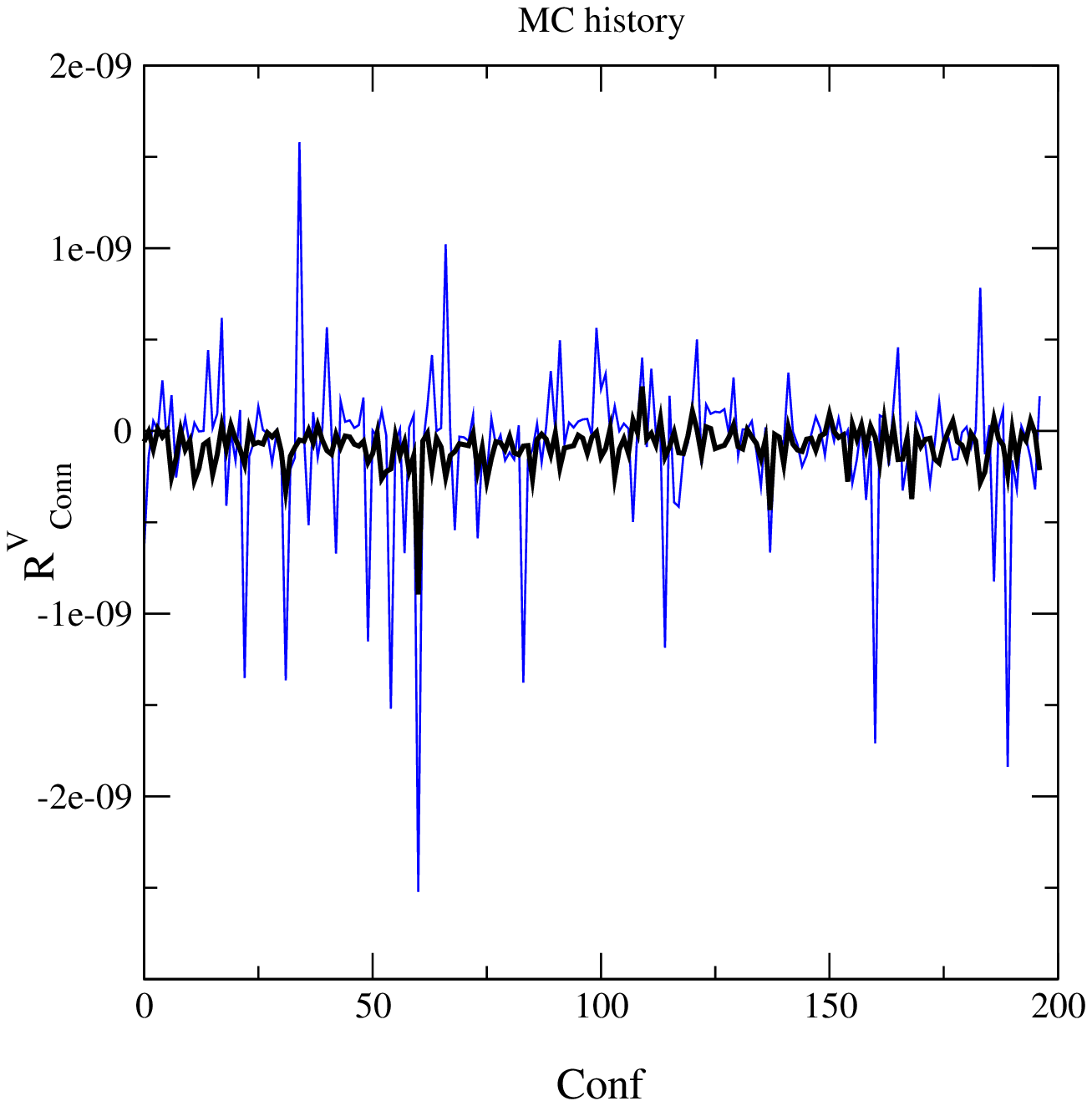}\hspace{0.5cm}\includegraphics[height=0.3\textheight,angle=-90]{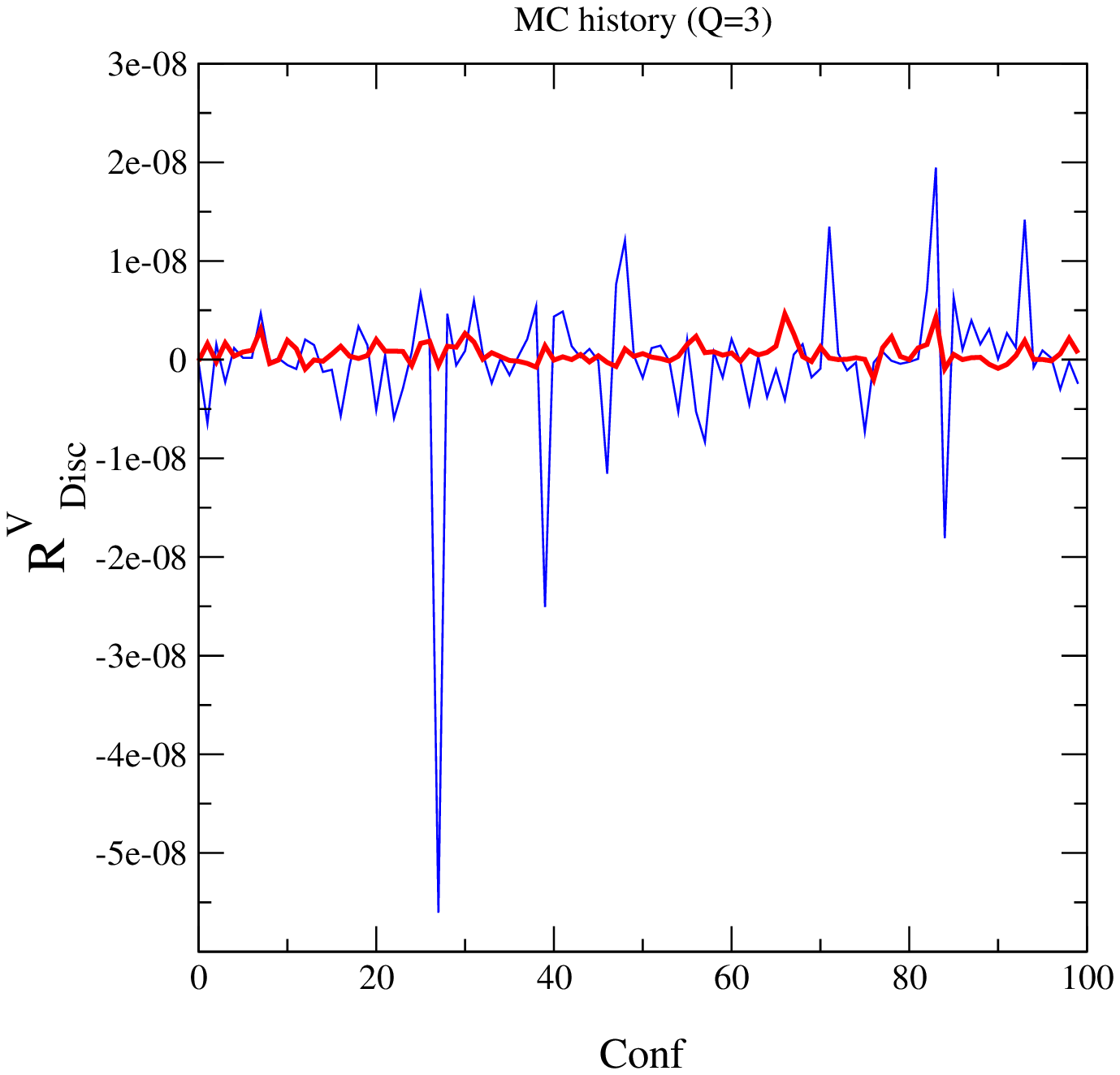} 
%\end{tabular}
\caption{Left: Montecarlo history with and without LMA of the color-connected contraction in $R^\sigma$ for the lightest mass in the $p$-regime. Right: Montecarlo history with and without LMA of the color-disconnected contraction of $R^\sigma_\nu$ in the $\epsilon$-regime for $|\nu|=3$. }
\label{fig:3ptlma}
\end{center}
\end{figure}

The LMA typically requires the computation of the low-lying eigenvalues and eigenfunctions and the inversion of the Dirac operator on them, which is a non-negligible overhead, but still pays off as 
Figure~\ref{fig:3ptlma} clearly shows. The two Montecarlo histories with and without LMA for one of the contractions of the three-point correlators of eq.(\ref{eq:3pt}) show that the improvement from LMA is very significant.

\begin{figure}
\begin{center}
%%%%%%%%%%%%%%%%%%%%%%%%%%%%%%%% FIGURE %%%%%%%%%%%%%%%%%%%%%%%%%%%%%%%%%%%%
\includegraphics[width=.45\textwidth]{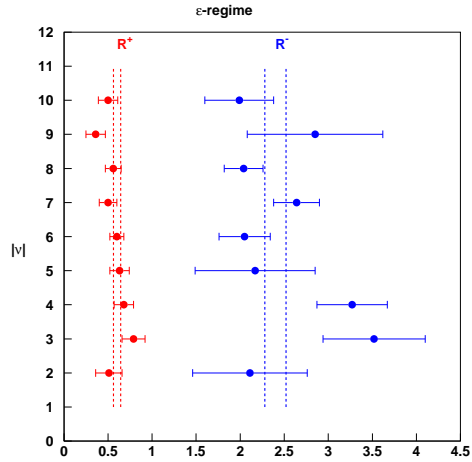}
\caption{Ratios $R^\pm$ as a function of the topological charge $2 \leq \nu \leq 10$.}
\label{fig:nudep}
%%%%%%%%%%%%%%%%%%%%%%%%%%%%%%%%%%%%%%%%%%%%%%%%%%%%%%%%%%%%%%%%%%%%%%%%%%%%
\end{center}
\end{figure}

\begin{figure}
\begin{center}
%%%%%%%%%%%%%%%%%%%%%%%%%%%%%%%% FIGURE %%%%%%%%%%%%%%%%%%%%%%%%%%%%%%%%%%%%
\includegraphics[width=.45\textwidth]{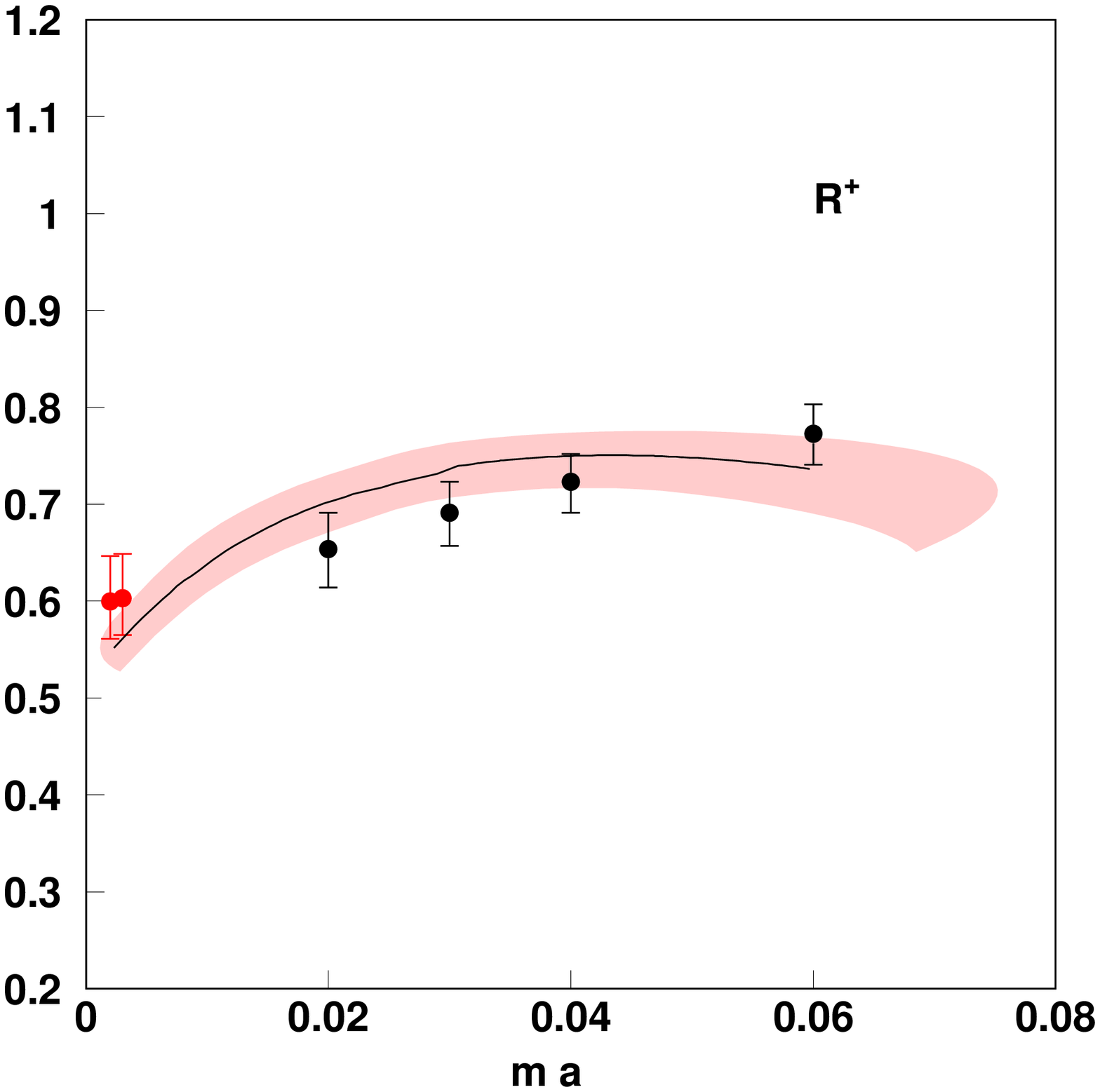} \includegraphics[width=.45\textwidth]{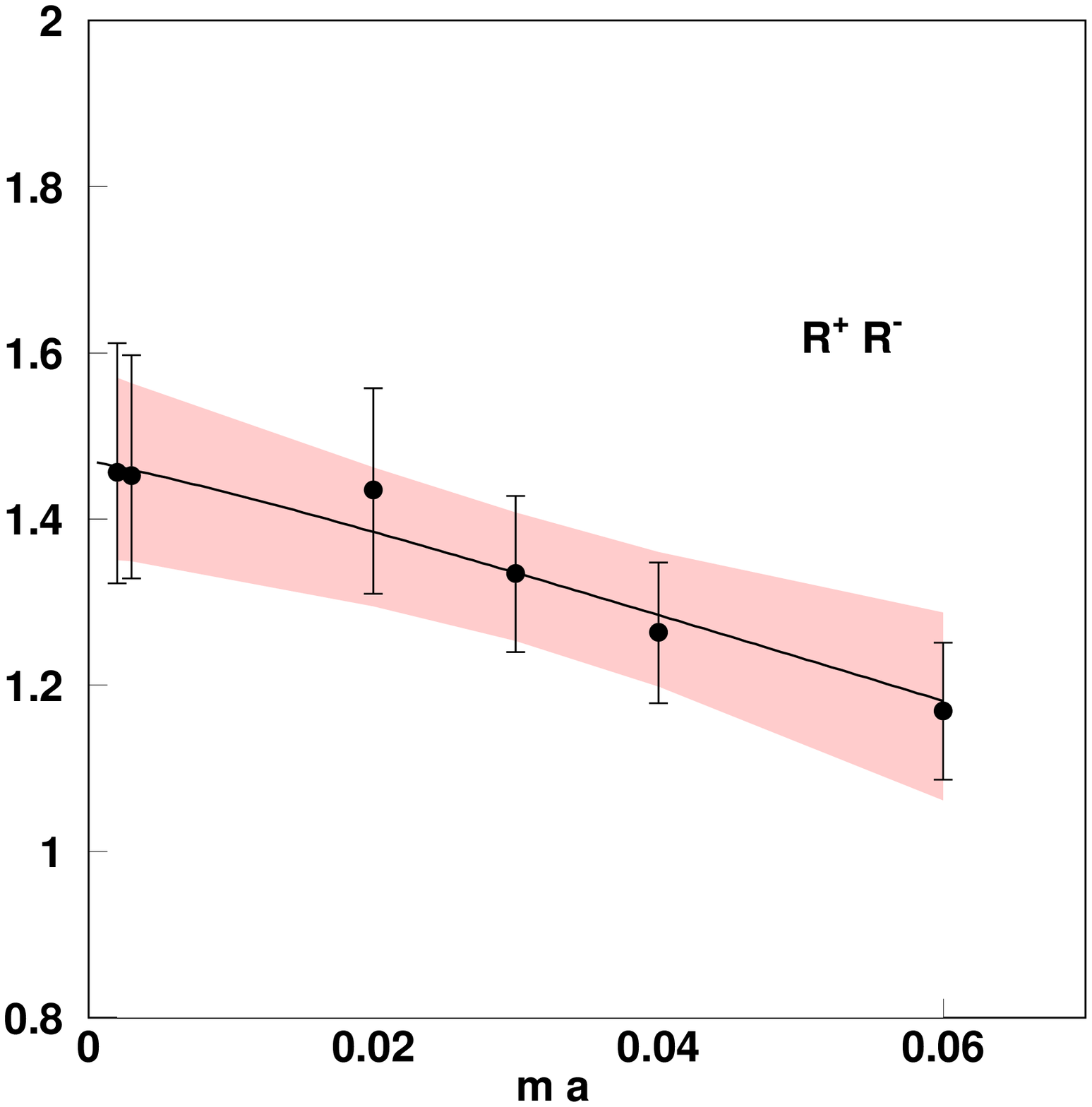}
\caption{Left: NLO chiral perturbation theory fits of the bare $R^+$ as a funcion of the quark mass. Right: NLO chiral perturbation theory fits of the bare
$R^+ R^-$ as a function of the quark mass. The bands represent the statistical 
errors on the fitting function. }
\label{fig:fits}
%%%%%%%%%%%%%%%%%%%%%%%%%%%%%%%%%%%%%%%%%%%%%%%%%%%%%%%%%%%%%%%%%%%%%%%%%%%%
\end{center}
\end{figure}

In \cite{prl} we have presented the first results of a lattice determination of $R^\sigma$ in the quenched approximation using overlap fermions \cite{overlap}. The simulation parameters are summarized in 
Table~\ref{tab:simul}. 
\begin{table}
\begin{center}
\begin{tabular}{cccccclr}
\hline
    & $\beta$ & $L/a$ & $T/a$ & $n_{\rm low}$ & $L[{\rm fm}]$ & $~~~m$ & \# cfgs \\
\hline
$\epsilon$-regime & 5.8485 & 16 & 32 & 20 & 2 & $m_s/40, m_s/60$ & $O(800)$ \\
$p$-regime & 5.8485 & 16 & 32 &20 & 2 & $m_2/2-m_s/6$ & $O(200)$ \\
\hline
\end{tabular}
\caption{Simulation parameters}
\label{tab:simul}
\end{center}
\end{table}
The expected features of $R^\sigma_\nu$ in the $\epsilon$-regime as predicted from chiral perturbation theory in eq.~(\ref{eq:k}) are well reproduced by the data. At large time separations, the
ratio shows a flat behaviour in $x_0$ and $y_0$. There is no signal of $\nu$ dependence in $R_\nu^\pm$ as shown in Figure~\ref{fig:nudep}. A weighted averaged
is used to combine the result for all topologies. 
Finally there is no visible dependence with the quark mass for the small values considered. On the other hand the quark mass dependence of the ratios $R^\pm$ in the $p$-regime is quite 
significant. Note that in the $p$-regime all topological sectors are averaged. In Figure~\ref{fig:fits} we show the results of the fits to NLO chiral perturbation theory expressions of the two combinations with smaller mass dependence: $R^+$ and 
the product $R^+ R^-$. The latter combination has the nice property that the NLO chiral corrections vanish at zero quark mass. Each of the fits has two free parameters $((g^+)_{bare},\Lambda_+)$ and $((g^+ g^-)_{bare}, \Lambda_{\pm})$, where $\Lambda_+,\Lambda_\pm$ is a combination of the NLO couplings that enter in the $p$-regime. The bands contain the statistical (bootstrap) errors. 
Combining the results obtained from the fits of the bare ratios with the Wilson coefficients and the non-perturbative renormalization constants we obtain our main result:
\ba
g^+ = 0.51(3)(5)(6), \qquad g^-= 2.6(1)(3)(3),
\ea
where the first error is statistical, the second comes from the renormalization factor and the third is a systematic error estimated from the dispersion of the values of $g^\pm$ obtained from different fiting strategies (for example fitting only the $\epsilon$-regime or only the $p$-regime points).  Although the results have been obtained for just one lattice spacing, scaling studies of several observables with overlap fermions have shown that scaling violations tend to be very small \cite{ww}. 

These numbers can be compared with those that would be obtained from experiment if the $\Delta I=3/2$ and $\Delta I=1/2$ amplitudes 
would be matched to the corresponding ones in LO chiral perturbation theory in the GIM limit:
\ba
g^+ \simeq 0.5, \qquad g^- \simeq 10.4 .
\ea
Therefore the value of $g_+$ is strikingly close to experiment already in the GIM limit, but the value of $g_-$ is a factor $\sim 4$ smaller. 
A significant enhancement is therefore observed in this limit
\ba
{A_0\over A_2} \simeq 6, 
\ea
which is not large enough to explain the experimental ratio but is already significant and cannot be abscribed to penguin operators nor penguin contractions.  

\section{Towards a heavier charm}

Once the LECS in the GIM limit are known, the charm quark can be increased. If the charm is still light so that charmed mesons can be treated as chiral degrees of freedom, 
it is possible to match the two chiral theories with and without the charmed mesons analytically, or in other words the charmed mesons can be integrated out of the $SU(4)$ chiral theory. 
In this way one recovers an $SU(3)$ chiral Lagrangian where the couplings $g_{27}$ and $g_8$ can be computed as functions of $g_\pm$ and the charm quark mass. Only when the charm quark mass
satisfies 
\ba
{m_u \Sigma \over F^2} \ll {m_c \Sigma \over F^2} \ll (4\pi F)^2 .
\ea
is this perturbative matching reliable. 

\begin{figure}
\begin{center}
%%%%%%%%%%%%%%%%%%%%%%%%%%%%%%%% FIGURE %%%%%%%%%%%%%%%%%%%%%%%%%%%%%%%%%%%%
\includegraphics[width=.45\textwidth]{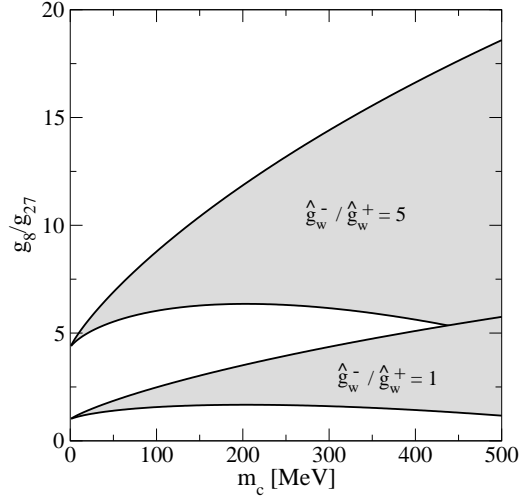}
\caption{Ratio of the low-energy couplings $g_8/g_{27}$ as a function of the charm quark mass for two different values of the GIM limit ratio $g^-/g^+ = 1$ and 5. The bands correspond to changing $\Lambda_\chi = 1-4~$GeV.}
\label{fig:largemc}
%%%%%%%%%%%%%%%%%%%%%%%%%%%%%%%%%%%%%%%%%%%%%%%%%%%%%%%%%%%%%%%%%%%%%%%%%%%%
\end{center}
\end{figure}

This exercise was carried out in Refs.~\cite{HerLai2, strat} with the result:
\ba
g_8(m_c)&=& {1\over 2} \biggl[ 
 {1\over 5} {g^+} \Bigl(1 + {15} \frac{M_c^2}{(4\pi F_{})^2}
                 \ln\frac{\Lambda_\chi}{M_c} \Bigr) +  {g^-} \Bigl( 1 + {3} \frac{M_c^2}{(4\pi F_{})^2}
                 \ln\frac{\Lambda_\chi}{M_c} \Bigr) \biggr]  \\
g_{27}(m_c)&=& {3 \over 5} g^+,  
\ea
where $\Lambda_{\chi}$ contains the unknown information on the NLO couplings that are relevant as $m_c$ increases. 
In Figure~\ref{fig:largemc} the dependence on $m_c$ of the ratio $g_8/g_{27}$ is shown for two values of the ratio of couplings in the GIM limit $g^-/g^+=1, 5$.  
The surprising observation is that only the octect coupling has a logarithmic 
enhancement. Unfortunately NLO couplings are needed to have predictability. 
The effect of these is represented by the band where they have been varied 
within reasonable values (the associated scale $\Lambda_{\chi} = 1-4$~GeV).
For a ratio of $g^-/g^+ \sim 5$, close to the value we have obtained in the GIM limit, there
could be a large effect coming from the integration of the charm. 
Unfortunately there is no much predictability unless the NLO couplings are known. 
In order to go beyond chiral perturbation theory it is necessary to do a 
non-perturbative matching this time with a heavy charm quark, which is the next step of our project.

It is well known that the case of a heavier charm will bring additional challenges. On the numerical side, the computation of three-point functions with a heavy charm requires the evaluation of the penguin contractions, which involve
a point-point propagator. We are confident that LMA will also help in this case, but this has yet to be demonstrated. On the theoretical side, the quenched ambiguities of Golterman and Pallante will be present in the $SU(3)$ chiral 
effective theory. It has been shown in \cite{HerLai3} that these quenching ambiguities are rather mild in the $\epsilon$-regime, and it is 
in principle possible to disentangle the ``physical'' couplings from the spurious ones.

\section{Conclusions}

The $\Delta I=1/2$ rule remains a big challenge for lattice QCD. We have presented a well-defined strategy to quantify 
the role of the charm quark mass and in particular of the penguin contractions in the enhancement \cite{strat}. The idea is to compute the low-energy
couplings mediating the $\Delta S=1$ transitions in a theory where the charm is light and degenerate with the remaining three quarks (GIM limit)
and compare them with those in a theory with just three light flavours, where the corresponding couplings  can be computed as a function of the 
charm quark mass. 

The GIM limit is easier to treat in many respects and the first results for the couplings have recently been presented in Ref.~\cite{prl} in the quenched approximation. The low-energy couplings in this limit already show a 
significant enhancement of the $\Delta I=1/2$ type. Even though the enhancement is not large enough to match the experiment, it already indicates that 
penguin operator/contractions cannot be the whole story. 

\section*{Acknowledgements}
I would like to thank my collaborators in the work presented here as well as in earlier work related to this subject: P.~Damgaard, M.C.~Diamantini, L.~Giusti, K.~Jansen, M.~Laine, L.~Lellouch, M.~L\"uscher,  C.~Pena, P.~Weisz, J. Wennekers and H. Wittig 
for an enjoyable collaboration and many illuminating discussions. I acknowledge partial support from the Spanish CICYT (Project No.\ FPA2004-00996 and FPA2005-01678) and  by the Generalitat Valenciana (Project No.\ GVA05/164).

\end{document}